\documentclass[epsfig,12pt]{article}
\usepackage[english]{babel} 
\usepackage{lmodern} 
\usepackage[utf8]{inputenc}
\usepackage[T1]{fontenc}

\usepackage{amssymb}
\usepackage{graphicx}
\usepackage{amsmath}
\usepackage{epsfig}
\usepackage[parfill]{parskip} 
\usepackage{wrapfig}
\usepackage{color}
\usepackage{vmargin}
\usepackage[most]{tcolorbox}
\usepackage[font={it}]{caption}

\usepackage{geometry}
\usepackage[round]{natbib}    
\setpapersize{A4}

\title{A Tracer-Based Algorithm for Automatic Generation of Seafloor Age Grids from Plate Tectonic Reconstructions}

\author{Krister S. Karlsen (\texttt{k.s.karlsen@geo.uio.no}), Mathew Domeier,\\Carmen Gaina and Clinton P. Conrad \\ \\ \large
Centre for Earth Evolution and Dynamics, University of Oslo, Norway}

\date{}

\begin{document}

\maketitle
\textbf{ \Large Abstract}

The age of the ocean floor and its time-dependent age distribution control fundamental features of the Earth, such as bathymetry, sea level and mantle heat loss. Recently, the development of increasingly sophisticated reconstructions of past plate motions has provided models for plate kinematics and plate boundary evolution back in geological time. These models implicitly include the information necessary to determine the age of ocean floor that has since been lost to subduction. However, due to the lack of an automated and efficient method for generating global seafloor age grids, many tectonic models, most notably those extending back into the Paleozoic, are published without an accompanying set of age models for oceanic lithosphere. Here we present an automatic, tracer-based algorithm that generates seafloor age grids from global plate tectonic reconstructions with defined plate boundaries. Our method enables us to produce the first seafloor age models for the Paleozoic's lost ocean basins. Estimated changes in sea level based on bathymetry inferred from our new age grids show good agreement with sea level record estimations from proxies, providing a possible explanation for the peak in sea level during the assembly phase of Pangea. This demonstrates how our seafloor age models can be directly compared with observables from the geologic record that extend further back in time than the constraints from preserved seafloor. Thus, our new algorithm may also aid the further development of plate tectonic reconstructions by strengthening the links between geological observations and tectonic reconstructions of deeper time.

\section{Introduction}

The discovery of a method to determine the age of the present-day oceanic crust, using reversals of the Earth's magnetic field \citep{vine1963magnetic}, gave rise to the recognition that the seafloor is spreading, and ultimately to the development and broad acceptance of plate tectonics. In the half-century since the plate tectonic revolution, detailed age models of the present-day oceanic lithosphere have been constructed, and digital global oceanic age grids are continuously refined \citep{muller1997digital,muller2008age,muller2016ocean}. A wealth of information, mainly from marine geophysical data, but also from geology of continental margins, were used to reconstruct the extent and age distribution of oceanic lithosphere of the past, including portions that have been subducted \citep{muller2008long}. These “paleo-seafloor age grids” present rich new opportunities for scientific inquiry, as a wide range of Earth processes can be further interrogated with the use of such age grids. Example applications include the estimation of paleobathymetry (spatial and temporal changes in ocean basin depth, which in turn is important for understanding past ocean currents and their effect on paleoclimate, e.g., \citealt{straume2019globsed}), sea level change \citep{muller2008long}, global seafloor heat flow \citep{loyd2007time, crameri2019dynamic}, and the subduction volume flux, which impacts geomagnetic reversals \citep{hounslow2018subduction}, the thermal structure of paleo-subduction zones \citep{maunder2019modeling}, water transport to the deep mantle \citep{karlsen2019deep}, and the slab pull force on tectonic plates \citep{conrad2004temporal,faccenna2012role}. Seafloor ages for past times are also important as a boundary condition for global mantle convection models \citep{gurnis2012plate}.

Present day age grid models are based on a set of isochrons (lines defined by equal seafloor ages) constructed using information from magnetic and gravity data available at various resolutions in most oceanic basins. Ages for seafloor locations between isochrons are computed based on rotation parameters that describe the plate motions for various time intervals. The isochrons and rotation parameters are linked to a specific geomagnetic timescale, and the choice of timescales will influence the calculated values of spreading velocities. To ensure a smooth grid of ocean floor ages that maintains sharp age discontinuities at fracture zones, \cite{muller1997digital} designed an algorithm where they first created a set of densely interpolated isochrons along plate flow lines, and then used a minimum curvature routine to obtain age values on a regular grid. This method for reconstructing seafloor age from present-day seafloor age data and a plate kinematic model is unfortunately extremely time-consuming and requires significant human input, and consequently may be subjective or introduce errors. Because of this, seafloor ages are usually only determined after a plate reconstruction model has been finalized; they have not previously been computed "on the fly" from the plate kinematic model itself.

The mid-ocean ridges constitute the locus of seafloor generation through time, while plate kinematics define the seafloor's subsequent journey until its destruction at a subduction zone. Thus, global plate tectonic reconstructions that define the motions of the plates and the locations and types of plate boundaries (for a detailed description of this type of model see \citealt{gurnis2012plate})  also implicitly define the age and structure of oceanic basins. Global plate tectonic models with dynamic plate boundaries have been constructed back into the mid-Paleozoic (410 Ma; \citealt{domeier2014plate,matthews2016global}), but published paleo-seafloor age grids are only available globally for the last 250 Ma \citep{muller2019global}. This timing discrepancy has, partly, occurred because the reconstruction of global paleo-seafloor age grids from a plate tectonic reconstruction presents a tedious and labor intensive task in the absence of an automatic method. Moreover, these reconstructions are subject to continuous changes as new geological information becomes available. It follows that an automatic and more efficient method for seafloor age determination is needed to allow the use of plate tectonic reconstructions to better decipher Earth's processes and dynamics through time. An automatic method also allows detecting inconsistencies in kinematic models, which would help to improve them. In this study we present such an automatic method for generating seafloor age grids, and introduce a specific implementation of it as an open-source Python code called \emph{Tracer Tectonics} (or \emph{TracTec}) \citep{karlsen2019tracer}.

\section{Methods}

We divide the computational approach for generating seafloor age grids from kinematic plate models into three modules: preprocessing, main algorithm and a post-processing (Fig. \ref{fig:flow_chart}). This division saves time by avoiding expensive data processing along the way, so that the main algorithm steps can remain as streamlined and efficient as possible.

\subsection{Preprocessing}

\begin{figure}[h!]
\vspace{0.5cm}
\makebox[\textwidth][c]{\includegraphics[width=1.15\textwidth]{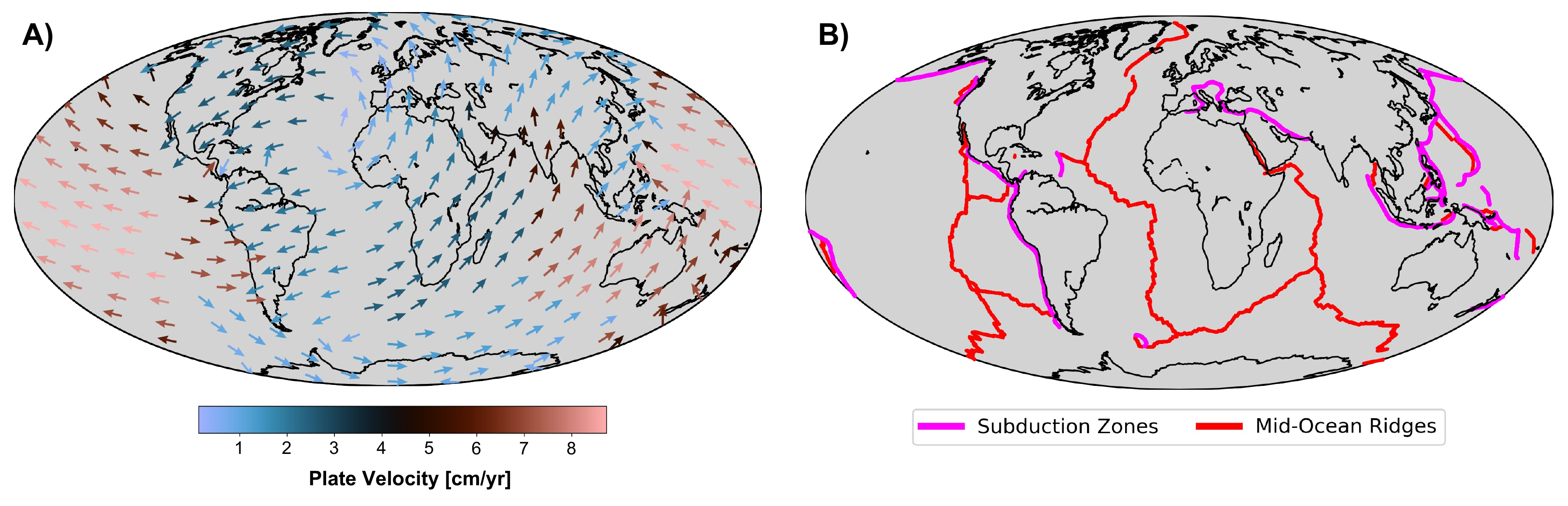}}
      \caption{Present day global surface velocity field (A) and locations of mid-ocean ridges and subduction zones (B) derived from global plate tectonic reconstructions \citep{matthews2016global}.}
      \label{fig:fig1}
      \vspace{0.5cm}
\end{figure}

\begin{figure}[h!]
\vspace{0.5cm}
\makebox[\textwidth][c]{\includegraphics[width=1.15\textwidth]{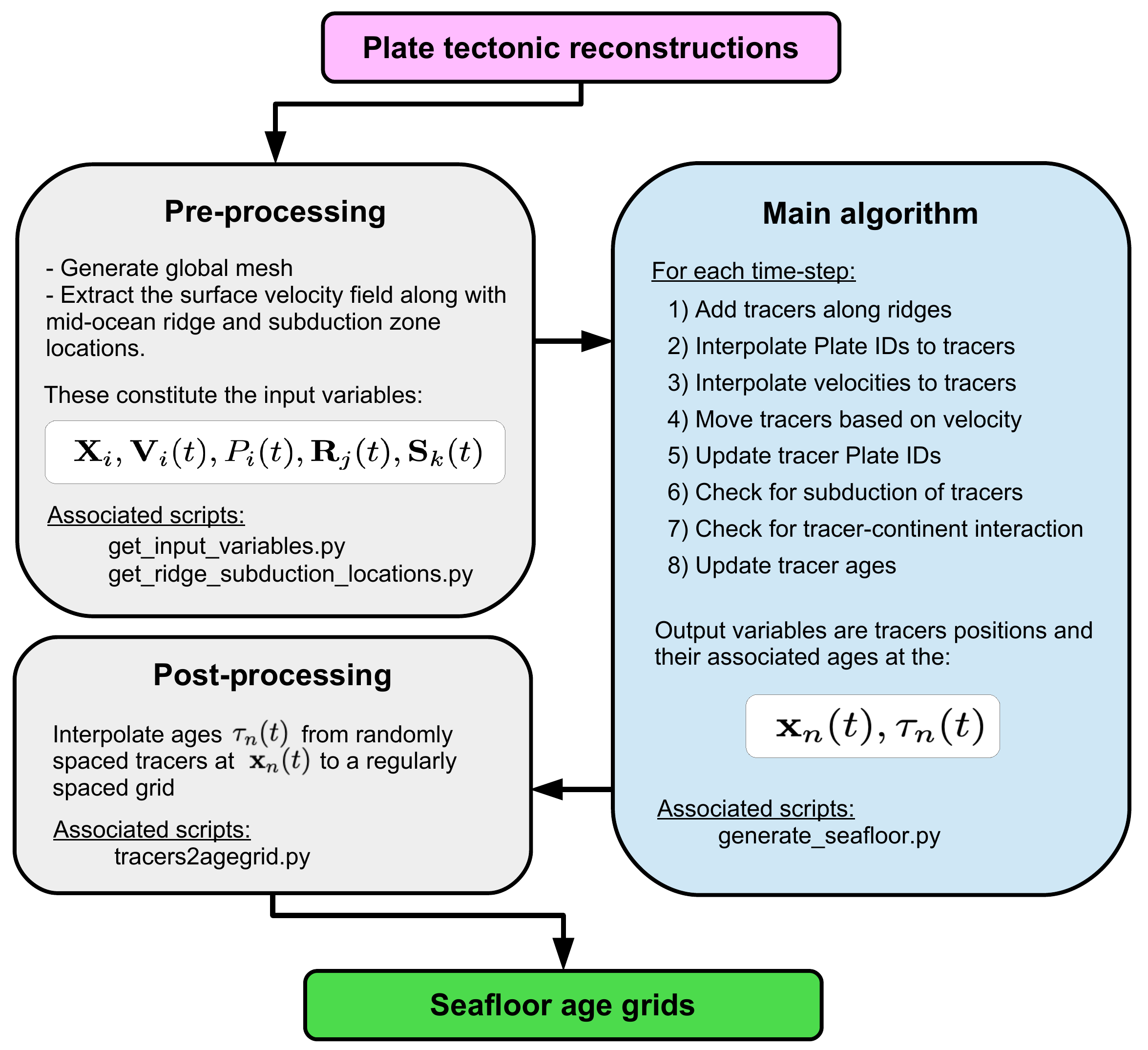}}
      \caption{Flow chart of TracTec \citep{karlsen2019tracer} showing an overview of the algorithm and the steps used to generate seafloor age grids from plate tectonic reconstructions. }
      \label{fig:flow_chart}
      \vspace{0.5cm}
\end{figure}

Global, full-plate tectonic reconstructions present a set of dynamically closed plate polygons that evolve over time from a set of dynamically evolving plate boundaries \citep{gurnis2012plate}, which are derived from a wealth of geological and geophysical data. Each plate is associated with a \emph{Plate ID} and its motion can be defined by rotation about an Euler pole, or an equivalent surface velocity field, at each time-step. The collection of all the plate velocity fields constitute a \emph{global surface velocity field} $\mathbf{V}(t)$  (Fig. \ref{fig:fig1}A) at each time-step $t \in [0, T]$, where $t = 0$ defines the present-day and $t = T$ is the earliest point in time for which the plate reconstruction is defined. The dynamically evolving plate boundaries are labelled according to the type of plate boundary that they represent; here we are interested in only those that are either a mid-ocean ridge or a subduction zone (Fig. \ref{fig:fig1}B).

As the time-dependent spatial distribution of mid-ocean ridges and subduction zones, together with plate kinematics, dictates the age of the ocean floor, we need to extract these properties from a given full-plate model and output them in a convenient format. To accomplish this, we use \emph{Pygplates}; a Python-based scripting interface to GPlates \citep{boyden2011next} that allows for easy automation of such tasks. To describe the plate kinematics, we define a global mesh with node coordinates given by $\mathbf{X}_i$, where $i = 0,1,..,N-1$, and $N$ is the number of mesh nodes. Thus, at some time $t$, a velocity field $\mathbf{V}_i(t)$ and a set of Plate IDs $P_i(t)$ can be assigned to each node of the mesh by interpolation from the full-plate model. Additionally, to be able to distinguish between oceanic and continental regions, we also extract the locations of continents through time and interpolate them to our globally defined mesh, such that each mesh node  is associated with a binary region value $C_i(t)$ through time. $C_i(t)=1$ indicates that this mesh node is associated with a continent, while $C_i(t)=0$ represents an ocean. Note that the mesh node coordinates $\mathbf{X}_i$ are static, while the properties (velocities $\mathbf{V}_i(t)$, Plate IDs $P_i(t)$  and region type $C_i(t)$) at these nodes change with time. There are no requirement on the type of mesh, for instance it could be a regular latitude-longitude mesh, or a CitcomS mesh \citep{zhong2000role}.

We extract plate boundaries labelled mid-ocean ridge and sample these circle arc segments at a user-defined resolution given by the parameter $\Delta_R$, defined the distance (km) between each point (or vertex). Because this resolution parameter also controls the number of tracers that are added at the mid-ocean ridges to track the evolution of the seafloor age (see next section), we recommend a relatively high resolution ($\Delta_R \sim 20-50 $ km). The collection of points with the 
desired spacing define mid-ocean ridge locations, and is denoted $\mathbf{R}_j(t)$, where $j = 0,1,..,N_R(t)-1$ and $N_R(t)$ is the number of ridge vertices at time $t$. Subduction zone plate boundary segments are extracted and sampled in the same manner, and the vertices that define their locations are given by $\mathbf{S}_k(t)$ for $k = 0,1,..,N_S(t)-1$, with the resolution given by $\Delta_S$ (values of $\sim 15-30 $ km are suggested).

We developed python-scripts that set up and extract the necessary input variables ($\mathbf{X}_i$,  $\mathbf{V}_i(t)$, $P_i(t)$, $\mathbf{R}_j(t)$  and $\mathbf{S}_k(t)$) to our algorithm from full-plate tectonic reconstructions. These can be found in: http://doi.org/10.5281/zenodo.3383154.

\subsection{Main Algorithm}

To simulate and track the journey of oceanic lithosphere through space and time; from its creation along a ridge until its destruction at a subduction zone, we use tracers. These are numerical particles on which quantities of interest are tracked. The essential property we want to study is the age of the oceanic lithosphere, and this is the primary property we track with the tracers. A secondary property we track is the Plate ID associated with each tracer, i.e. the plate to which each tracer currently belongs. We track Plate IDs to determine when a tracer crosses a plate boundary (recognized by a change in its associated Plate ID), which in turn is useful to detect subduction of tracers. At a given time $t$, the number of tracers is $N_T(t)$, their positions are $\mathbf{x}_n(t)$, their age $\tau_n(t)$, and their associated Plate IDs are $p_n(t)$, for $n = 0,1,..,N_T(t)-1$.

We break the algorithm into individual sub-steps (see below, 1-8) that are completed at every time-step $t \in [0, T]$, before moving on to the next:

\noindent 
\emph{1) Add tracers} \\ 
At the beginning of each time-step we add tracers on each side of the mid-ocean ridges, with an offset distance of 30 km away from the ridge axis to each side. This is to ensure that tracers get assigned the correct plate velocity from the global velocity field. To demonstrate the importance of this strategy we provide an example: the lowermost ridge-axis in Figure 3 A-C lines up parallel to the mesh nodes. If tracers were added exactly (zero offset distance) along the ridge-axis, all would be assigned a velocity to the left, and a sparse region would open up as they were moved. The spacing between the added tracers (on each side of the ridge) is given by $\Delta_R$, and the age $\tau_n (t)$ of these tracers  is set to one (this is equivalent of assuming an average spearing rate of 3 cm/yr for the first one Myr).

\noindent 
\emph{2) Interpolate Plate ID to tracers} \\ 
Plate IDs $P_i(t)$ are interpolated from the mesh nodes $\mathbf{X}_i$ to the tracer positions $\mathbf{x}_n(t)$ using the N-dimensional nearest neighbor algorithm from the Python library Scipy \citep{jones2001scipy}, obtaining $p_n(t)$. 

\begin{figure}[h!]
\vspace{0.5cm}
\makebox[\textwidth][c]{\includegraphics[width=1.1\textwidth]{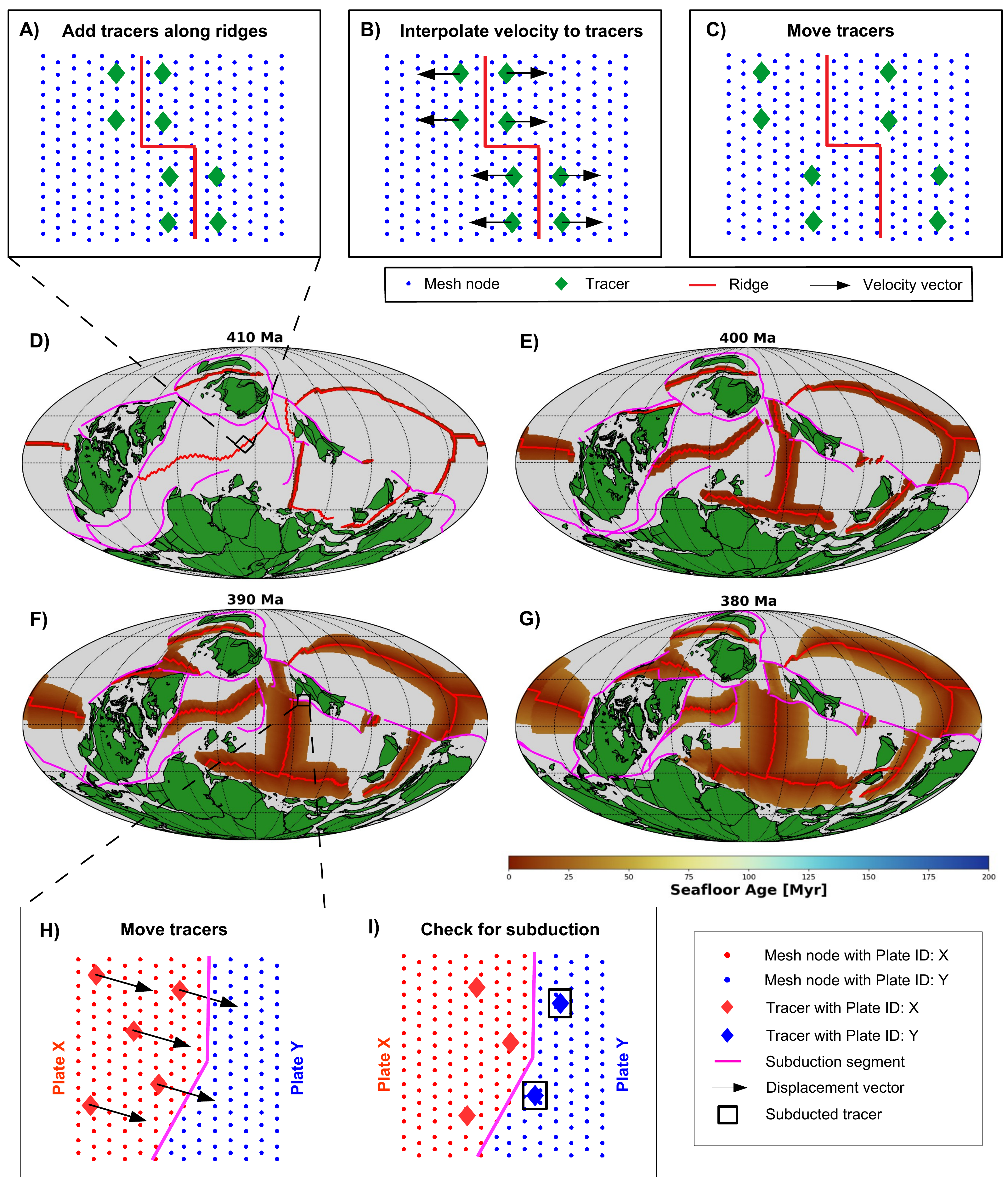}}
      \caption{Figure 3: Schematics illustrating how tracers are added along mid-ocean ridges (A), moved (B-C), and checked for subduction (H-I). The repeated application of these steps over time leads to an ocean basin gradually filled with tracers that track the age of the ocean floor (D-G). In this example the model of \cite{matthews2016global} is used. Note that (A-C, H-I) are not to scale, thus density of tracers and mesh nodes shown here are not representative.}
      \label{fig:mod_evo}
      \vspace{0.5cm}
\end{figure}

\noindent 
\emph{3) Interpolate velocity to tracers} \\ 
The tracers that track the oceanic lithosphere's journey must move according to the plate velocities known only at the mesh nodes $\mathbf{X}_i$. Therefore $\mathbf{V}_i(t)$ must be interpolated to the tracer positions $\mathbf{x}_n(t)$ (Fig. \ref{fig:mod_evo}B). We employ the same nearest neighbor algorithm for the velocities, because this conserves the discontinuous nature of tectonic plate velocities and avoids unphysical smoothing across plate boundaries, while providing superior efficiency on a sphere. The interpolated velocities at $\mathbf{x}_n(t)$ are denoted $\mathbf{v}_n(t)$.

\noindent 
\emph{4) Move tracers} \\ 
To obtain the next positions of the tracers (Fig. \ref{fig:mod_evo} B-C) at the time-step $t+\Delta t$, we use the forward-Euler method: $\mathbf{x}_n(t+\Delta t) = \mathbf{x}_n(t) + \Delta t\mathbf{v}_n(t)$. The default time-step $\Delta t$ is set to 1 Myr, which is appropriately small in the context of average plate speeds. Notably, although our algorithm is perfectly suited to operate at any arbitrarily shorter (or larger) time-steps, the temporal resolution of most full-plate models is $\sim$ 1 Myr. Thus, use of shorter time-steps can result in erroneous behavior owing to shortcomings in the temporal resolution of the input model.

\noindent 
\emph{5) Update Plate IDs} \\ 
Plate IDs $P_i(t+\Delta t)$ are interpolated to the new tracer positions $\mathbf{x}_n(t+\Delta t) $, obtaining $p_n(t+\Delta t)$.

\noindent 
\emph{6) Check for subduction} \\ 
After having moved the tracers from $\mathbf{x}_n(t) $ to $\mathbf{x}_n(t+\Delta t) $, we check if any tracers have been subducted (Fig. \ref{fig:mod_evo} H-I). This is accomplished by first determining which tracers have changed Plate ID by comparing $p_n(t)$ against $p_n(t+\Delta t)$. Given the sub-set of tracers that have changed Plate ID, we determine if any are within distance $R_{min}$ of a subduction zone. The default value of $R_{min}$ is set to 100 km, which we have found to be appropriate. Any tracers that fulfill both of these criteria are considered subducted and are thus terminated at this time-step.

\noindent 
By default, the subducted tracers are stored in auxiliary files that can be used for studying various properties of the subduction history, for example to estimate the average age of subducting plates through time, or to consider the time-integrated history of subduction.

\noindent 
\emph{7) Check for tracer-continent interaction} \\ 
To ensure that tracers don't end up on continents (which shouldn't happen in closed-polygon models, but is not impossible in nature (e.g., ophiolites) and can occur due to inexorable flaws in full-plate models), we delete all tracers whose closest mesh node $\mathbf{X}_i$ is associated with a continent $C_i(t)$. From the perspective of computational efficiency, it is also preferred to maintain only the tracers that are strictly needed. 

\noindent 
\emph{8) Update tracer ages} \\ 
Before moving on to the next time-step, which practically means returning to step 1) of the algorithm, we update the age of the tracers that are left after the two filtering steps 6-7), by simply by adding $\Delta t$ to their current age, obtaining $\tau_n (t+\Delta t)$.

Note that as pygplates (and the GPlates files associated with the input plate model) are used only in preprocessing step, the age grid algorithm (steps 1-8) could be applied to output from a global mantle convection model, in the case the user had a way of tracking the distribution of plates and their associated boundaries and velocities at the surface (e.g. \citealt{mallard2017adopt}).

\subsection{Post-Processing}

The output from steps 1-8) is randomly distributed tracer positions $\mathbf{x}_n(t)$ associated with ages $\tau_n (t)$. However, for most applications, it is convenient to express the seafloor ages on a regular grid, rather than at random points. This calls for a post-processing step that interpolates seafloor ages from tracer positions onto a regular grid. There are countless ways of doing this, ranging from simple nearest neighbor algorithms, to weighted means, to splines etc. Depending on the application, smoothing of resulting agegrids may be preferred or required. In our online repository we provide an example of a simple post-processing script that uses GMT's linear interpolation algorithm \citep{wessel2013generic} to obtain seafloor ages on a regular grid.

The Python scripts to generate age grids based on the steps described above using Tracer Tectonics is provided in the Zenodo repository \citep{karlsen2019tracer}. A summary and a general overview of the algorithm are shown in Figure 2.

\subsection{Initial Condition}

Running the algorithm without any initial condition, as in the example of Figure 3 D-G, we see that it takes some tens of millions of years before the ocean basins are entirely covered by tracers. Technically, this is the time it takes for the predicted seafloor ages to result solely from the plate kinematics of the input model. Alternatively, one could apply an initial condition that incorporates some educated guess of the seafloor ages for the initial time step. This is straightforward to include in the framework presented (steps 1-8 of the main algorithm), by simply initializing $\mathbf{x}_n(t=0)$ and $\tau_n (t=0)$. As with any time-dependent model, one should be aware of the assumptions implicit to the chosen initial condition, and its effects on the model output. For this algorithm (and in contrast to many geodynamic models), it is straightforward to track the effect of an applied initial condition. This can be achieved by simply tracking the fraction of initial tracers through time (Fig. S1); at some point, all the initial tracers will be eliminated, from which point the output will no longer depend on the initial condition.

\section{Discussion}

\subsection{Validation and Benchmarking}

\begin{figure}[h!]
\vspace{0.5cm}
\makebox[\textwidth][c]{\includegraphics[width=1.1\textwidth]{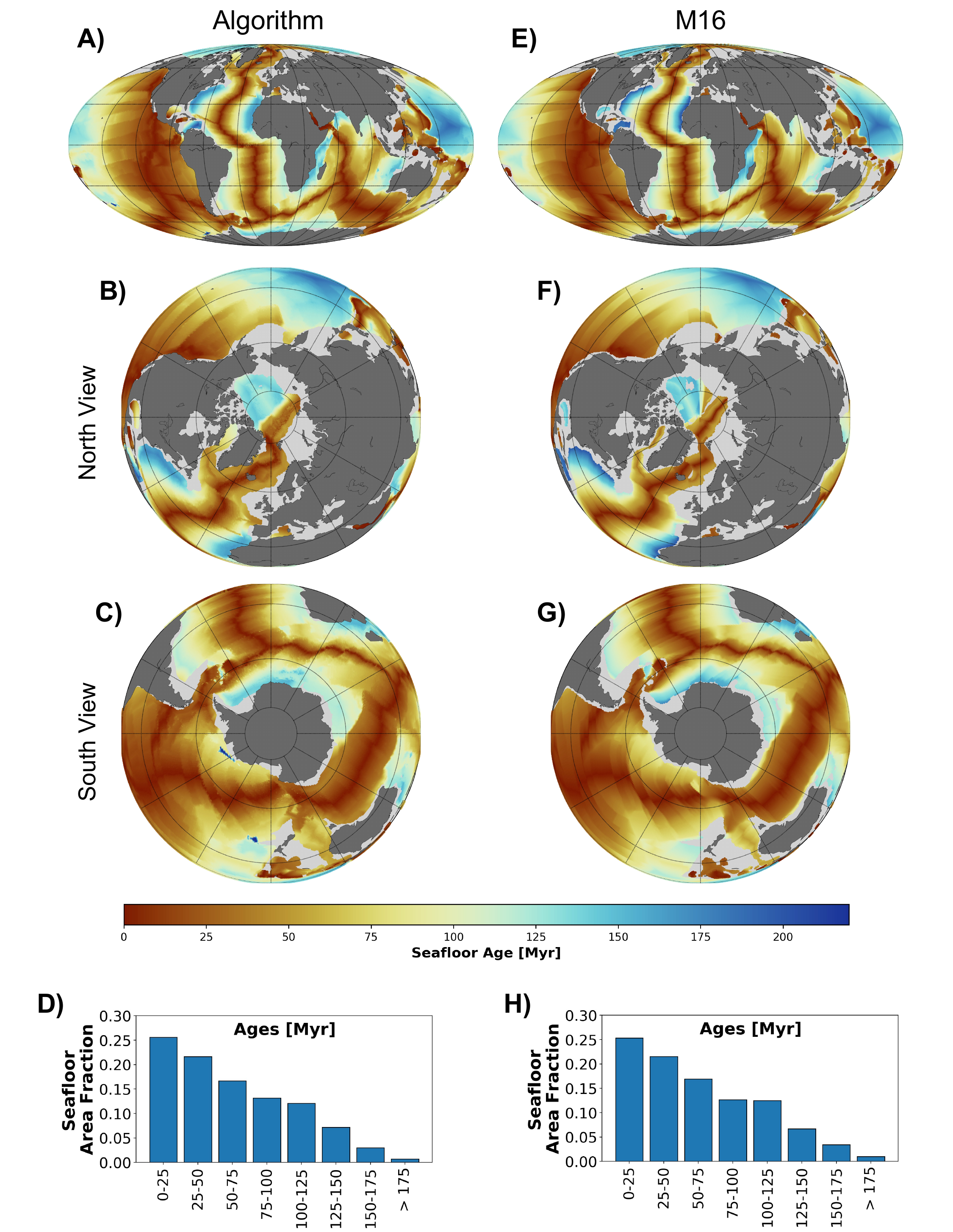}}
      \caption{Maps comparing present-day seafloor ages generated by our algorithm (A-C) to those of M16 \citep{muller2016ocean} model (E-G), and their corresponding age-area distributions (D and H). The input plate model used here was \citep{matthews2016global}.}
      \label{fig:pd_bm}
      \vspace{0.5cm}
\end{figure}

\begin{figure}[h!]
\vspace{0.5cm}
\makebox[\textwidth][c]{\includegraphics[width=1.2\textwidth]{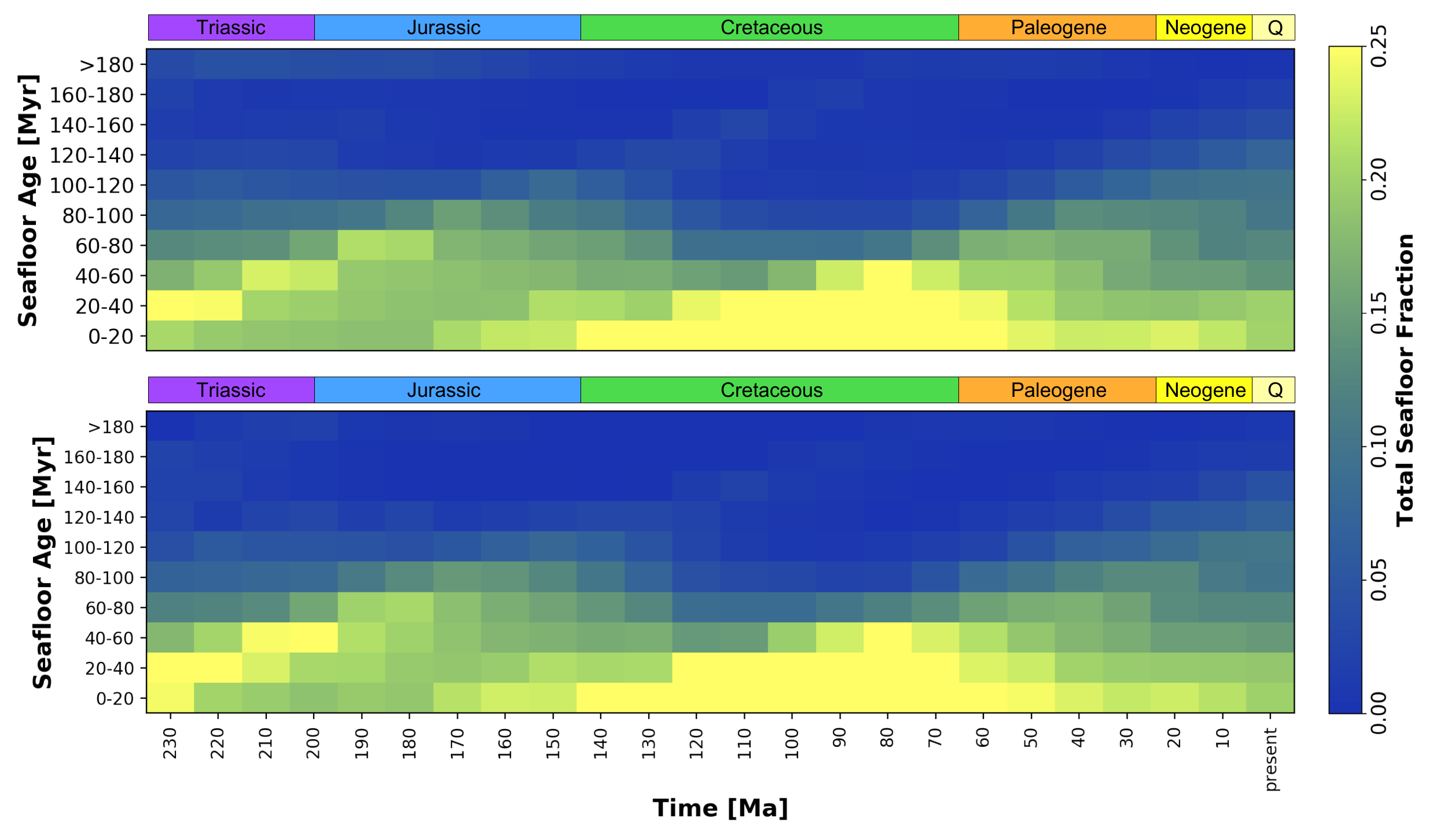}}
      \caption{Comparison of the age-area distribution of the seafloor through time for our algorithm created age grids (top) and the \citep{muller2016ocean} (M16) model (bottom).}
      \label{fig:agedist}
      \vspace{0.5cm}
\end{figure}

For computational methods to effectively contribute to advances in the field of geoscience (as well as in the natural sciences in general), output from algorithms and simulations should be validated against relevant geological model constraints. This is particularly true when introducing new modeling approaches and software, making the practice of code benchmarking an essential and necessary part of software development. To validate our algorithm and its implementation, we compare the seafloor age grids generated by our algorithm against published present-day seafloor age models (Fig. \ref{fig:pd_bm}), direct point observations of present-day seafloor ages (Fig. S2) and against the time-dependent age-area distribution of the oceanic lithosphere for the last 230 Myr (Fig. \ref{fig:agedist}) based on M16 \citep{muller2016ocean}.

\newpage

\clearpage
\newpage

We used \cite{matthews2016global} as the input plate model, with the initial condition shown in Figure \ref{fig:sea_level}A, to generate the seafloor age grids used to evaluate the performance of our algorithm in this section.

To generate the seafloor age grids used to evaluate the performance of our algorithm in this section, we employed \cite{matthews2016global} as the input plate model with the initial condition shown in Figure \ref{fig:sea_level}A. Notably, however, the effect of an initial condition at 400 Ma is very small already by 300 Ma, and zero at the present-day (Fig. S1). The initial condition and the age grids are available from our online repository \citep{karlsen2019tracer}.

Our algorithm reproduces the present-day seafloor well (Fig. \ref{fig:pd_bm} A-C), as can be seen from the maps comparing the resulting age grids to those of M16 (Fig. \ref{fig:pd_bm} E-G). The characteristic triangular present-day seafloor age-area distribution (\citealt{sclater1980heat, cogne2004temporal}), which shows the fraction of the ocean floor that falls within a certain age range, is also well reproduced. On the regional scale, some minor differences can be seen, for example there are three narrow bands of artificially young seafloor branching from the Mid-Atlantic Ridge near the Caribbean and the south-west of Iberia. These are merely consequences of the underlying plate model \citep{matthews2016global}, for which these plate boundaries are erroneously designated as mid-ocean ridges, either at present (Fig. 1), or in the recent past. This demonstrates that a combined work-flow for developing plate tectonic reconstructions, which includes analysis of the seafloor ages, can reveal flaws and inconsistencies in the model. In the case of the aforementioned errors in the mid-Atlantic, simply re-labeling the offending boundaries as transform features and re-running the age grid algorithm addresses the issue.

To further evaluate the performance of our algorithm, we compare the generated present-day age grid against ages inferred from magnetic reversal picks (Fig. S2). This global dataset provides by far the most comprehensive, direct sources of oceanic lithosphere ages, and is available from the Global Seafloor Fabric and Magnetic Lineation (GSFML) database \citep{seton2014community}. We see that about 30 \% of the 101418 pick ages are within $\pm 1$ Myr of the ages from our age grids, and about 80 \% are within $\pm 5$ Myr (Fig. S2B). These numbers are slightly higher for models of the present-day ocean floor that are based on more labor-intensive methods (described in Section 1) such as Muller et al. (2008b), for which they are $\sim$ 45 \% and $\sim$ 92 \% respectively (Fig. S2C), and $\sim$ 52 \% and $\sim$ 93 \% for M16 (Fig. S2D). 

As pointed out in several studies (e.g.  \citealt{becker2009past,coltice2012dynamic,sim2016influence}), the triangular age-area distribution of the present-day ocean floor is unlikely to be a constant feature through Earth’s history. In particular, large fluctuations are predicted to have occurred in the rates of seafloor spreading and global subduction, (e.g.   \citealt{domeier2018episodic,karlsen2019deep}), which should preclude a constant age-area distribution of the seafloor. Moreover, the age-area distribution of the seafloor is an important feature of our planet because it exhibits a first-order control on e.g. bathymetry, sea level, and planetary cooling through regulation of surface heat flow (i.e. loss of mantle heat). Therefore, as a third benchmark, we compare the time-dependent age-area distributions of the seafloor as generated by our algorithm with those published in \cite{muller2016ocean}.

\clearpage

We observe a clear time-dependence in the age-area distributions computed from our age grids, and this time-dependence is nearly identical to that predicted by M16 (Fig. \ref{fig:agedist}). Given the broad observed trend toward relatively younger seafloor ages between ~140-80 Ma, we anticipate that some of these seafloor age variations are related to the widespread development of new ridge systems during the Cretaceous. The initiation of some of these new ridge systems (including the southern mid-Atlantic) was associated with late-stage Pangea breakup, whereas other new ridges appeared in the Paleo-Pacific basin, probably related to the emplacement of large igneous provinces \citep{torsvik2019pacific}. The rapid development of these new ridges, producing juvenile oceanic lithosphere, must have been synchronously balanced with increased subduction that would have accelerated the destruction of relatively old seafloor; and these two processes combined thus explains the tendency toward relatively younger seafloor during the Cretaceous (Fig. \ref{fig:agedist}).

In summary, our algorithm reproduces detailed reconstructions of seafloor ages like those of M16. We observe only minor regional differences between the algorithm-generated present-day seafloor age grid and the M16 model. These differences mainly occur in regions where \cite{matthews2016global} inferred plate boundary locations that deviate from the interpreted isochrons of M16. This merely demonstrates that our algorithm is an automatic and fast way to detect inconsistencies between data and models. Global features like the age-area distribution, which is important for many geodynamic applications, are robustly reproduced over time. Thus, as plate tectonic reconstructions improve, so will the reliability and regional resolution of the age-grids produced by our algorithm.

\subsection{An Example Application: Paleo Sea Level}

In this section we will demonstrate how our new algorithm and paleo-age grids (computed as described in Section 3.1) can be used to study tectonic mechanisms for sea level variations during the last 400 Myr. We would like to stress that the age-grids generated by our algorithm are bound to the input plate model, and as is the case for all plate models operating in times earlier than 200 Ma, the kinematics of all oceanic plates are necessarily synthetic (because no in situ oceanic lithosphere older than 200 Ma has survived to the present-day). This naturally implies that the construction of age-grids for earlier times is much more uncertain, and interpretation of them should be done with caution and care; here we only use these synthetic age grids to consider some global, first-order trends for the sake of demonstration.

\begin{figure}[h!]
\vspace{0.5cm}
\makebox[\textwidth][c]{\includegraphics[width=1.15\textwidth]{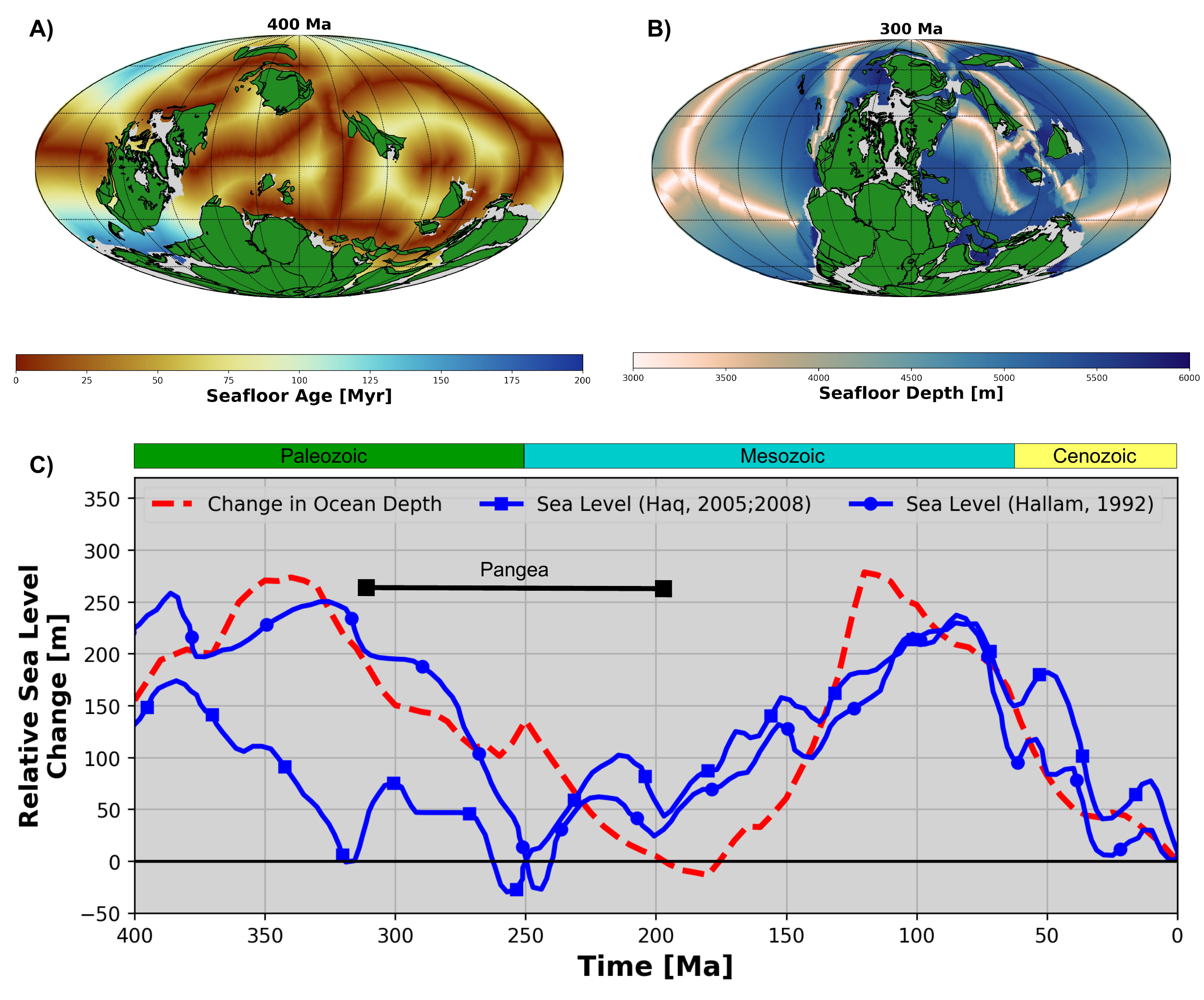}}
      \caption{Initial condition (A) used to generate seafloor age grids from the plate reconstruction of \cite{matthews2016global} from 400 Ma to the present, and associated bathymetry (B) at 300 Ma, computed by applying \cite{crosby2009analysis}. From the bathymetry models, we compute changes in mean ocean depth and isostatically compensate them \citep{pitman1978relationship} to compare sea level changes relative to the present day (C, red dashed line) to Phanerozoic sea level reconstructions (C, blue lines) inferred from the sedimentary record \citep{hallam1992phanerozoic,haq2005phanerozoic,haq2008chronology}. Thick black bar shows the approximate duration of Pangea.}
      \label{fig:sea_level}
      \vspace{0.5cm}
\end{figure}

From the generated seafloor age grids we compute bathymetry by applying the age-depth relation of \citep{crosby2009analysis}. Next, we use these bathymetry grids (e.g., Fig. \ref{fig:sea_level}B) to compute the change in average ocean basin depth relative to the present-day. Finally, we compare how these changes in average ocean depth would affect sea level, and relate them to the sea level history (Fig. \ref{fig:sea_level}C) as inferred from the sedimentary record \citep{hallam1992phanerozoic,haq2005phanerozoic,haq2008chronology}. Although many other processes affect sea level fluctuations on tectonic time scales, the age-area distribution of the seafloor (through thermal subsidence of the ageing oceanic lithosphere) exhibits the first-order control, and is the only process that shows a direct correlation with the sea level record \citep{muller2008long,conrad2013solid,karlsen2019deep}. Therefore, an automatic method for generating seafloor age grids (from which ocean depth can be computed) enables sea level to be used as a first-order order deep-time constraint on plate tectonic reconstructions.

Our prediction of sea level fluctuations caused by ocean depth changes inferred from our 400 Myr reconstruction of age grids shows a first-order agreement with established sea level records \citep{hallam1992phanerozoic,haq2005phanerozoic,haq2008chronology}. Such a correlation has been noted previously into the Mesozoic (e.g., \citealt{muller2008long,conrad2013solid}), but our algorithm applied to the plate reconstructions of Matthews et al. 2016 shows that this correlation may extend back to the late Paleozoic. Our age models predict a clear peak in sea level during the late stages of Pangea assembly (Fig. \ref{fig:sea_level}C), in agreement with the early predictions of sea-level change based on conceptual models of a supercontinental cycle \citep{worsley1985proterozoic,worsley1986tectonic,nance1986post}. Moreover, the consistency between predicted and observed sea level indicates that the tectonic rates (seafloor spreading and its counterpart, seafloor subduction) in the underlying plate tectonic model might be reasonable for this period of time.

\subsection{Limitations}

The uncertainties in the generated seafloor age grids are directly linked to, and controlled by, the underlying plate tectonic model. Thus, the generated age-grids are dependent on global plate velocities and the locations of mid-ocean ridges and subduction zones, for which the assignment of formal errors is either impractical or impossible. For these reasons, we expect uncertainty in the age grids to be proportional to those of the underlying plate tectonic reconstructions, with negligible additional uncertainty introduced through the rather straightforward computational steps of our algorithm (Fig. 2). A final word of caution is that the generated age grids will never be better than the input plate model. We thus advise users to familiarize themselves with the uncertainties tied to the underlying plate tectonic reconstructions.

\clearpage

\section{Conclusions}

The development of plate tectonic reconstructions in the past decade has been rapid, with the introduction of powerful new tools \citep{boyden2011next,gurnis2012plate,shephard2017consistency} and the augmentation of global kinematic datasets \citep{seton2014community,gaina2019global} propelling a concomitant proliferation of new and emergent full-plate models, of both regional
\citep{shephard2013tectonic,zahirovic2014cretaceous,
gibbons2015tectonic,domeier2016plate,zahirovic2016tectonic,
domeier2018early,torsvik2019pacific} and global scope 
\citep{seton2012global,domeier2014plate,muller2016ocean,matthews2016global,merdith2017full,muller2019global}. Part of the power of these latest full-plate models is that they implicitly provide information on plate kinematics across the entire global (or regional) surface, and so can be used to derive seafloor age grids to test and evaluate them, as well as to potentially explore other geophysical questions that relate to seafloor ages. Unfortunately, while the information needed to compute seafloor ages is implicitly available, the presently-established workflows to retrieve and process this information are time-consuming, laborious and not publicly available. This emphasizes the need for a method that can automatically generate seafloor age grids from full-plate tectonic reconstructions. 

In this study we have presented an algorithm for generating seafloor age grids that robustly predicts the known present-day ocean floor ages, as well as a previously published model for the time-dependent age-area distribution. This new method produced the first set of oceanic lithosphere age grids that approximate the first order age distribution of the Paleozoic oceans. Application of our generated seafloor models to estimate past sea level changes reveals a general agreement with observations from the sedimentary stratigraphy record \citep{hallam1992phanerozoic,haq2005phanerozoic,haq2008chronology}, and provides a possible explanation for a peak in sea level during the assembly phase of Pangea. We hope that our automated algorithm will enable such comparisons between age-grid predictions and existing geological constraints to become a routine procedure of the full-plate model development process. Such an improved workflow should ultimately lead to better and more self-consistent combined reconstructions of both plate tectonics and seafloor ages.

\section*{Computer Code Availability}

\noindent
\underline{Name of code:} Tracer Tectonics (TracTec) \\
\underline{Developer:} Krister S. Karlsen  \\
\underline{First available:} September 1, 2019. \\
\underline{Hardware required:} No requirements.\\
\underline{Program language:} Python. \\
\underline{Software dependencies:} GMT (only post-processing) and the following Python libraries: pygplates (only preprocessing), scipy, numpy.  \\
\underline{Program size:} 860 MB (including example input files, benchmark age grids, etc.)\\
\underline{Source code:} https://doi.org/10.5281/zenodo.3383154

\section*{Acknowledgments}

This paper has benefited from discussions with members of the CEED Earth Modelling group. All calculations in the main algorithm were preformed with Python \citep{van2011python}, while GMT \citep{wessel2013generic} was used for post-processing. We thank Fabio Crameri for the development of perceptually uniform and color-vision-deficiency friendly color maps \citep{crameri2018scientific1}, ensuring an accurate representation of the underlying data in the figures \citep{crameri2018geodynamic}.

This research was funded by the Research Council of Norway's Centres of Excellence Project 223272.

\bibliography{mybibfile}

\begin{thebibliography}{53}
\providecommand{\natexlab}[1]{#1}
\providecommand{\url}[1]{\texttt{#1}}
\expandafter\ifx\csname urlstyle\endcsname\relax
  \providecommand{\doi}[1]{doi: #1}\else
  \providecommand{\doi}{doi: \begingroup \urlstyle{rm}\Url}\fi

\bibitem[Becker et~al.(2009)Becker, Conrad, Buffett, and
  M{\"u}ller]{becker2009past}
Becker, Thorsten~W; Conrad, Clinton~P; Buffett, Bruce, and M{\"u}ller,
  R~Dietmar.
\newblock Past and present seafloor age distributions and the temporal
  evolution of plate tectonic heat transport.
\newblock \emph{Earth and Planetary Science Letters}, 278\penalty0
  (3-4):\penalty0 233--242, 2009.

\bibitem[Boyden et~al.(2011)Boyden, M{\"u}ller, Gurnis, Torsvik, Clark, Turner,
  Ivey-Law, Watson, and Cannon]{boyden2011next}
Boyden, James~A; M{\"u}ller, R~Dietmar; Gurnis, Michael; Torsvik, Trond~H;
  Clark, James~A; Turner, Mark; Ivey-Law, Hamish; Watson, Robin~J, and Cannon,
  John~S.
\newblock Next-generation plate-tectonic reconstructions using gplates.
\newblock 2011.

\bibitem[Cogn{\'e} and Humler(2004)]{cogne2004temporal}
Cogn{\'e}, Jean-Pascal and Humler, Eric.
\newblock Temporal variation of oceanic spreading and crustal production rates
  during the last 180 my.
\newblock \emph{Earth and Planetary Science Letters}, 227\penalty0
  (3-4):\penalty0 427--439, 2004.

\bibitem[Coltice et~al.(2012)Coltice, Rolf, Tackley, and
  Labrosse]{coltice2012dynamic}
Coltice, Nicolas; Rolf, Tobias; Tackley, Paul~J, and Labrosse, St{\'e}phane.
\newblock Dynamic causes of the relation between area and age of the ocean
  floor.
\newblock \emph{Science}, 336\penalty0 (6079):\penalty0 335--338, 2012.

\bibitem[Conrad(2013)]{conrad2013solid}
Conrad, Clinton~P.
\newblock The solid earth’s influence on sea level.
\newblock \emph{GSA Bulletin}, 125\penalty0 (7-8):\penalty0 1027--1052, 2013.

\bibitem[Conrad and Lithgow-Bertelloni(2004)]{conrad2004temporal}
Conrad, Clinton~P and Lithgow-Bertelloni, Carolina.
\newblock The temporal evolution of plate driving forces: Importance of “slab
  suction” versus “slab pull” during the cenozoic.
\newblock \emph{Journal of Geophysical Research: Solid Earth}, 109\penalty0
  (B10), 2004.

\bibitem[Crameri(2018{\natexlab{a}})]{crameri2018geodynamic}
Crameri, Fabio.
\newblock Geodynamic diagnostics, scientific visualisation and staglab 3.0.
\newblock \emph{Geoscientific Model Development}, 11\penalty0 (6):\penalty0
  2541--2562, 2018{\natexlab{a}}.

\bibitem[Crameri(2018{\natexlab{b}})]{crameri2018scientific1}
Crameri, Fabio.
\newblock Scientific colour-maps.
\newblock Zenodo. http://doi.org/10.5281/zenodo.1243862, 2018{\natexlab{b}}.

\bibitem[Crameri et~al.(2019)Crameri, Conrad, Mont{\'e}si, and
  Lithgow-Bertelloni]{crameri2019dynamic}
Crameri, Fabio; Conrad, Clinton~P; Mont{\'e}si, Laurent, and
  Lithgow-Bertelloni, Carolina~R.
\newblock The dynamic life of an oceanic plate.
\newblock \emph{Tectonophysics}, 760:\penalty0 107--135, 2019.

\bibitem[Crosby and McKenzie(2009)]{crosby2009analysis}
Crosby, AG and McKenzie, D.
\newblock An analysis of young ocean depth, gravity and global residual
  topography.
\newblock \emph{Geophysical Journal International}, 178\penalty0 (3):\penalty0
  1198--1219, 2009.

\bibitem[Domeier(2016)]{domeier2016plate}
Domeier, Mathew.
\newblock A plate tectonic scenario for the iapetus and rheic oceans.
\newblock \emph{Gondwana Research}, 36:\penalty0 275--295, 2016.

\bibitem[Domeier(2018)]{domeier2018early}
Domeier, Mathew.
\newblock Early paleozoic tectonics of asia: Towards a full-plate model.
\newblock \emph{Geoscience Frontiers}, 9\penalty0 (3):\penalty0 789--862, 2018.

\bibitem[Domeier and Torsvik(2014)]{domeier2014plate}
Domeier, Mathew and Torsvik, Trond~H.
\newblock Plate tectonics in the late paleozoic.
\newblock \emph{Geoscience Frontiers}, 5\penalty0 (3):\penalty0 303--350, 2014.

\bibitem[Domeier et~al.(2018)Domeier, Magni, Hounslow, and
  Torsvik]{domeier2018episodic}
Domeier, Mathew; Magni, Valentina; Hounslow, Mark~W, and Torsvik, Trond~H.
\newblock Episodic zircon age spectra mimic fluctuations in subduction.
\newblock \emph{Scientific reports}, 8\penalty0 (1):\penalty0 17471, 2018.

\bibitem[Faccenna et~al.(2012)Faccenna, Becker, Lallemand, and
  Steinberger]{faccenna2012role}
Faccenna, Claudio; Becker, Thorsten~W; Lallemand, Serge, and Steinberger,
  Bernhard.
\newblock On the role of slab pull in the cenozoic motion of the pacific plate.
\newblock \emph{Geophysical Research Letters}, 39\penalty0 (3), 2012.

\bibitem[Gaina and Jakob(2019)]{gaina2019global}
Gaina, Carmen and Jakob, Johannes.
\newblock Global eocene tectonic unrest: Possible causes and effects around the
  north american plate.
\newblock \emph{Tectonophysics}, 760:\penalty0 136--151, 2019.

\bibitem[Gibbons et~al.(2015)Gibbons, Zahirovic, M{\"u}ller, Whittaker, and
  Yatheesh]{gibbons2015tectonic}
Gibbons, AD; Zahirovic, S; M{\"u}ller, RD; Whittaker, JM, and Yatheesh, V.
\newblock A tectonic model reconciling evidence for the collisions between
  india, eurasia and intra-oceanic arcs of the central-eastern tethys.
\newblock \emph{Gondwana Research}, 28\penalty0 (2):\penalty0 451--492, 2015.

\bibitem[Gurnis et~al.(2012)Gurnis, Turner, Zahirovic, DiCaprio, Spasojevic,
  M{\"u}ller, Boyden, Seton, Manea, and Bower]{gurnis2012plate}
Gurnis, Michael; Turner, Mark; Zahirovic, Sabin; DiCaprio, Lydia; Spasojevic,
  Sonja; M{\"u}ller, R~Dietmar; Boyden, James; Seton, Maria; Manea,
  Vlad~Constantin, and Bower, Dan~J.
\newblock Plate tectonic reconstructions with continuously closing plates.
\newblock \emph{Computers \& Geosciences}, 38\penalty0 (1):\penalty0 35--42,
  2012.

\bibitem[Hallam(1992)]{hallam1992phanerozoic}
Hallam, Anthony.
\newblock \emph{Phanerozoic sea-level changes}.
\newblock Columbia University Press, 1992.

\bibitem[Haq and Al-Qahtani(2005)]{haq2005phanerozoic}
Haq, Bilal~U and Al-Qahtani, Abdul~Motaleb.
\newblock Phanerozoic cycles of sea-level change on the arabian platform.
\newblock \emph{GeoArabia}, 10\penalty0 (2):\penalty0 127--160, 2005.

\bibitem[Haq and Schutter(2008)]{haq2008chronology}
Haq, Bilal~U and Schutter, Stephen~R.
\newblock A chronology of paleozoic sea-level changes.
\newblock \emph{Science}, 322\penalty0 (5898):\penalty0 64--68, 2008.

\bibitem[Hounslow et~al.(2018)Hounslow, Domeier, and
  Biggin]{hounslow2018subduction}
Hounslow, Mark~W; Domeier, Mathew, and Biggin, Andrew~J.
\newblock Subduction flux modulates the geomagnetic polarity reversal rate.
\newblock \emph{Tectonophysics}, 742:\penalty0 34--49, 2018.

\bibitem[Jones et~al.(2001)Jones, Oliphant, Peterson, et~al.]{jones2001scipy}
Jones, Eric; Oliphant, Travis; Peterson, Pearu, and others, .
\newblock {SciPy}: Open source scientific tools for {Python}, 2001.
\newblock URL \url{http://www.scipy.org/}.

\bibitem[Karlsen et~al.(2019{\natexlab{a}})Karlsen, Conrad, and
  Magni]{karlsen2019deep}
Karlsen, Krister~S; Conrad, Clinton~P, and Magni, Valentina.
\newblock Deep water cycling and sea level change since the breakup of pangea.
\newblock \emph{Geochemistry, Geophysics, Geosystems}, 20\penalty0
  (6):\penalty0 2919--2935, 2019{\natexlab{a}}.

\bibitem[Karlsen et~al.(2019{\natexlab{b}})Karlsen, Domeier, Gaina, and
  Conrad]{karlsen2019tracer}
Karlsen, Krister~S; Domeier, Mathew; Gaina, Carmen, and Conrad, Clinton~P.
\newblock {TracerTectonics} (version 1.0).
\newblock Zenodo, 2019{\natexlab{b}}.
\newblock URL \url{http://doi.org/10.5281/zenodo.3383154}.

\bibitem[Loyd et~al.(2007)Loyd, Becker, Conrad, Lithgow-Bertelloni, and
  Corsetti]{loyd2007time}
Loyd, SJ; Becker, TW; Conrad, CP; Lithgow-Bertelloni, C, and Corsetti, FA.
\newblock Time variability in cenozoic reconstructions of mantle heat flow:
  plate tectonic cycles and implications for earth's thermal evolution.
\newblock \emph{Proceedings of the National Academy of Sciences}, 104\penalty0
  (36):\penalty0 14266--14271, 2007.

\bibitem[Mallard et~al.(2017)Mallard, Jacquet, and Coltice]{mallard2017adopt}
Mallard, C; Jacquet, B, and Coltice, N.
\newblock Adopt: A tool for automatic detection of tectonic plates at the
  surface of convection models.
\newblock \emph{Geochemistry, Geophysics, Geosystems}, 18\penalty0
  (8):\penalty0 3197--3208, 2017.

\bibitem[Matthews et~al.(2016)Matthews, Maloney, Zahirovic, Williams, Seton,
  and Mueller]{matthews2016global}
Matthews, Kara~J; Maloney, Kayla~T; Zahirovic, Sabin; Williams, Simon~E; Seton,
  Maria, and Mueller, R~Dietmar.
\newblock Global plate boundary evolution and kinematics since the late
  paleozoic.
\newblock \emph{Global and Planetary Change}, 146:\penalty0 226--250, 2016.

\bibitem[Maunder et~al.(2019)Maunder, van Hunen, Bouilhol, and
  Magni]{maunder2019modeling}
Maunder, B; van Hunen, J; Bouilhol, P, and Magni, V.
\newblock Modeling slab temperature: A reevaluation of the thermal parameter.
\newblock \emph{Geochemistry, Geophysics, Geosystems}, 20\penalty0
  (2):\penalty0 673--687, 2019.

\bibitem[Merdith et~al.(2017)Merdith, Collins, Williams, Pisarevsky, Foden,
  Archibald, Blades, Alessio, Armistead, Plavsa, et~al.]{merdith2017full}
Merdith, Andrew~S; Collins, Alan~S; Williams, Simon~E; Pisarevsky, Sergei;
  Foden, John~D; Archibald, Donnelly~B; Blades, Morgan~L; Alessio, Brandon~L;
  Armistead, Sheree; Plavsa, Diana, and others, .
\newblock A full-plate global reconstruction of the neoproterozoic.
\newblock \emph{Gondwana Research}, 50:\penalty0 84--134, 2017.

\bibitem[M{\"u}ller et~al.(1997)M{\"u}ller, Roest, Royer, Gahagan, and
  Sclater]{muller1997digital}
M{\"u}ller, R~Dietmar; Roest, Walter~R; Royer, Jean-Yves; Gahagan, Lisa~M, and
  Sclater, John~G.
\newblock Digital isochrons of the world's ocean floor.
\newblock \emph{Journal of Geophysical Research: Solid Earth}, 102\penalty0
  (B2):\penalty0 3211--3214, 1997.

\bibitem[M{\"u}ller et~al.(2008{\natexlab{a}})M{\"u}ller, Sdrolias, Gaina, and
  Roest]{muller2008age}
M{\"u}ller, R~Dietmar; Sdrolias, Maria; Gaina, Carmen, and Roest, Walter~R.
\newblock Age, spreading rates, and spreading asymmetry of the world's ocean
  crust.
\newblock \emph{Geochemistry, Geophysics, Geosystems}, 9\penalty0 (4),
  2008{\natexlab{a}}.

\bibitem[M{\"u}ller et~al.(2008{\natexlab{b}})M{\"u}ller, Sdrolias, Gaina,
  Steinberger, and Heine]{muller2008long}
M{\"u}ller, R~Dietmar; Sdrolias, Maria; Gaina, Carmen; Steinberger, Bernhard,
  and Heine, Christian.
\newblock Long-term sea-level fluctuations driven by ocean basin dynamics.
\newblock \emph{science}, 319\penalty0 (5868):\penalty0 1357--1362,
  2008{\natexlab{b}}.

\bibitem[M{\"u}ller et~al.(2016)M{\"u}ller, Seton, Zahirovic, Williams,
  Matthews, Wright, Shephard, Maloney, Barnett-Moore, Hosseinpour,
  et~al.]{muller2016ocean}
M{\"u}ller, R~Dietmar; Seton, Maria; Zahirovic, Sabin; Williams, Simon~E;
  Matthews, Kara~J; Wright, Nicky~M; Shephard, Grace~E; Maloney, Kayla~T;
  Barnett-Moore, Nicholas; Hosseinpour, Maral, and others, .
\newblock Ocean basin evolution and global-scale plate reorganization events
  since pangea breakup.
\newblock \emph{Annual Review of Earth and Planetary Sciences}, 44:\penalty0
  107--138, 2016.

\bibitem[M{\"u}ller et~al.(2019)M{\"u}ller, Zahirovic, Williams, Cannon, Seton,
  Bower, Tetley, Heine, Le~Breton, Liu, et~al.]{muller2019global}
M{\"u}ller, R~Dietmar; Zahirovic, Sabin; Williams, Simon~E; Cannon, John;
  Seton, Maria; Bower, Dan~J; Tetley, Michael~G; Heine, Christian; Le~Breton,
  Eline; Liu, Shaofeng, and others, .
\newblock A global plate model including lithospheric deformation along major
  rifts and orogens since the triassic.
\newblock \emph{Tectonics}, 2019.

\bibitem[Nance et~al.(1986)Nance, Worsley, and Moody]{nance1986post}
Nance, R~Damian; Worsley, Thomas~R, and Moody, Judith~B.
\newblock Post-archean biogeochemical cycles and long-term episodicity in
  tectonic processes.
\newblock \emph{Geology}, 14\penalty0 (6):\penalty0 514--518, 1986.

\bibitem[Pitman~III(1978)]{pitman1978relationship}
Pitman~III, Walter~C.
\newblock Relationship between eustacy and stratigraphic sequences of passive
  margins.
\newblock \emph{Geological Society of America Bulletin}, 89\penalty0
  (9):\penalty0 1389--1403, 1978.

\bibitem[Sclater et~al.(1980)Sclater, Jaupart, and Galson]{sclater1980heat}
Sclater, JjG; Jaupart, Cr, and Galson, D\_.
\newblock The heat flow through oceanic and continental crust and the heat loss
  of the earth.
\newblock \emph{Reviews of Geophysics}, 18\penalty0 (1):\penalty0 269--311,
  1980.

\bibitem[Seton et~al.(2012)Seton, M{\"u}ller, Zahirovic, Gaina, Torsvik,
  Shephard, Talsma, Gurnis, Turner, Maus, et~al.]{seton2012global}
Seton, M; M{\"u}ller, RD; Zahirovic, S; Gaina, C; Torsvik, T; Shephard, G;
  Talsma, A; Gurnis, M; Turner, M; Maus, S, and others, .
\newblock Global continental and ocean basin reconstructions since 200 ma.
\newblock \emph{Earth-Science Reviews}, 113\penalty0 (3-4):\penalty0 212--270,
  2012.

\bibitem[Seton et~al.(2014)Seton, Whittaker, Wessel, M{\"u}ller, DeMets,
  Merkouriev, Cande, Gaina, Eagles, Granot, et~al.]{seton2014community}
Seton, Maria; Whittaker, Joanne~M; Wessel, Paul; M{\"u}ller, R~Dietmar; DeMets,
  Charles; Merkouriev, Sergey; Cande, Steve; Gaina, Carmen; Eagles, Graeme;
  Granot, Roi, and others, .
\newblock Community infrastructure and repository for marine magnetic
  identifications.
\newblock \emph{Geochemistry, Geophysics, Geosystems}, 15\penalty0
  (4):\penalty0 1629--1641, 2014.

\bibitem[Shephard et~al.(2013)Shephard, M{\"u}ller, and
  Seton]{shephard2013tectonic}
Shephard, Grace~E; M{\"u}ller, R~Dietmar, and Seton, Maria.
\newblock The tectonic evolution of the arctic since pangea breakup:
  Integrating constraints from surface geology and geophysics with mantle
  structure.
\newblock \emph{Earth-Science Reviews}, 124:\penalty0 148--183, 2013.

\bibitem[Shephard et~al.(2017)Shephard, Matthews, Hosseini, and
  Domeier]{shephard2017consistency}
Shephard, Grace~Elizabeth; Matthews, Kara~J; Hosseini, Kasra, and Domeier,
  Mathew.
\newblock On the consistency of seismically imaged lower mantle slabs.
\newblock \emph{Scientific reports}, 7\penalty0 (1):\penalty0 10976, 2017.

\bibitem[Sim et~al.(2016)Sim, Stegman, and Coltice]{sim2016influence}
Sim, Shi~J; Stegman, Dave~R, and Coltice, Nicolas.
\newblock Influence of continental growth on mid-ocean ridge depth.
\newblock \emph{Geochemistry, Geophysics, Geosystems}, 17\penalty0
  (11):\penalty0 4425--4437, 2016.

\bibitem[Straume et~al.(2019)Straume, Gaina, Medvedev, Hochmuth, Gohl,
  Whittaker, Abdul~Fattah, Doornenbal, and Hopper]{straume2019globsed}
Straume, EO; Gaina, Carmen; Medvedev, S; Hochmuth, Katharina; Gohl, Karsten;
  Whittaker, Joanne~M; Abdul~Fattah, R; Doornenbal, JC, and Hopper, John~R.
\newblock Globsed: Updated total sediment thickness in the world's oceans.
\newblock \emph{Geochemistry, Geophysics, Geosystems}, 20\penalty0
  (4):\penalty0 1756--1772, 2019.

\bibitem[Torsvik et~al.(2019)Torsvik, Steinberger, Shephard, Doubrovine, Gaina,
  Domeier, Conrad, and Sager]{torsvik2019pacific}
Torsvik, Trond~H; Steinberger, Bernhard; Shephard, Grace~E; Doubrovine,
  Pavel~V; Gaina, Carmen; Domeier, Mathew; Conrad, Clinton~P, and Sager,
  William~W.
\newblock Pacific-panthalassic reconstructions: Overview, errata and the way
  forward.
\newblock \emph{Geochemistry, Geophysics, Geosystems}, 20\penalty0
  (7):\penalty0 3659--3689, 2019.

\bibitem[Van~Rossum and Drake(2011)]{van2011python}
Van~Rossum, Guido and Drake, Fred~L.
\newblock \emph{The python language reference manual}.
\newblock Network Theory Ltd., 2011.

\bibitem[Vine and Matthews(1963)]{vine1963magnetic}
Vine, Fred~J and Matthews, Drummond~Hoyle.
\newblock Magnetic anomalies over oceanic ridges.
\newblock \emph{Nature}, 199\penalty0 (4897):\penalty0 947--949, 1963.

\bibitem[Wessel et~al.(2013)Wessel, Smith, Scharroo, Luis, and
  Wobbe]{wessel2013generic}
Wessel, Paul; Smith, Walter~HF; Scharroo, Remko; Luis, Joaquim, and Wobbe,
  Florian.
\newblock Generic mapping tools: improved version released.
\newblock \emph{Eos, Transactions American Geophysical Union}, 94\penalty0
  (45):\penalty0 409--410, 2013.

\bibitem[Worsley et~al.(1986)Worsley, Nance, and Moody]{worsley1986tectonic}
Worsley, Thomas~R; Nance, R~Damian, and Moody, Judith~B.
\newblock Tectonic cycles and the history of the earth's biogeochemical and
  paleoceanographic record.
\newblock \emph{Paleoceanography}, 1\penalty0 (3):\penalty0 233--263, 1986.

\bibitem[Worsley et~al.(1985)Worsley, Moody, and Nance]{worsley1985proterozoic}
Worsley, TR; Moody, JB, and Nance, RD.
\newblock Proterozoic to recent tectonic tuning of biogeochemical cycles.
\newblock \emph{The carbon cycle and atmospheric CO2: natural variations
  Archean to present}, 32:\penalty0 561--572, 1985.

\bibitem[Zahirovic et~al.(2014)Zahirovic, Seton, and
  M{\"u}ller]{zahirovic2014cretaceous}
Zahirovic, S; Seton, M, and M{\"u}ller, RD.
\newblock The cretaceous and cenozoic tectonic evolution of southeast asia.
\newblock \emph{Solid Earth}, 5\penalty0 (1):\penalty0 227, 2014.

\bibitem[Zahirovic et~al.(2016)Zahirovic, Matthews, Flament, M{\"u}ller, Hill,
  Seton, and Gurnis]{zahirovic2016tectonic}
Zahirovic, Sabin; Matthews, Kara~J; Flament, Nicolas; M{\"u}ller, R~Dietmar;
  Hill, Kevin~C; Seton, Maria, and Gurnis, Michael.
\newblock Tectonic evolution and deep mantle structure of the eastern tethys
  since the latest jurassic.
\newblock \emph{Earth-Science Reviews}, 162:\penalty0 293--337, 2016.

\bibitem[Zhong et~al.(2000)Zhong, Zuber, Moresi, and Gurnis]{zhong2000role}
Zhong, Shijie; Zuber, Maria~T; Moresi, Louis, and Gurnis, Michael.
\newblock Role of temperature-dependent viscosity and surface plates in
  spherical shell models of mantle convection.
\newblock \emph{Journal of Geophysical Research: Solid Earth}, 105\penalty0
  (B5):\penalty0 11063--11082, 2000.

\end{thebibliography}

\newpage
\includegraphics[scale=0.75,page=1]{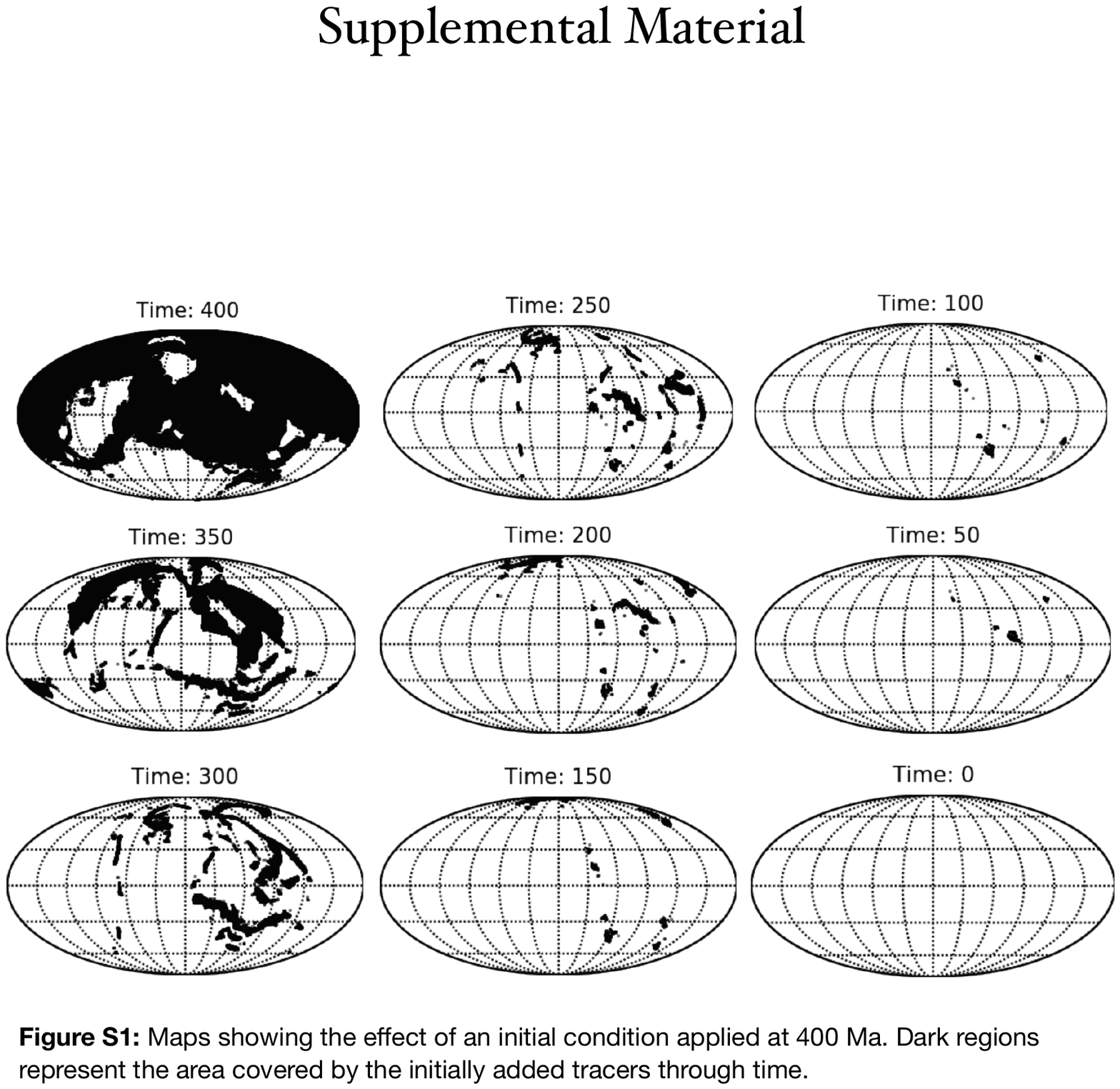}
\newpage
\includegraphics[scale=0.75,page=2]{supplemental_materials.pdf}

\end{document}